\documentclass[journal]{IEEEtran}
\ifCLASSINFOpdf
  \usepackage[pdftex]{graphicx}
  \graphicspath{{./figs/}{../jpeg/}}
   \DeclareGraphicsExtensions{.pdf,.jpeg,.png}
\else
  \usepackage[dvips]{graphicx}
  \graphicspath{{./figs/}}  
  \DeclareGraphicsExtensions{.eps}
\fi

\usepackage[cmex10]{amsmath}
\usepackage{algpseudocode}

\usepackage{array}

\usepackage{float}
\usepackage{amssymb}
\usepackage{multirow}
\usepackage{booktabs}
\usepackage{mdwmath}
\usepackage{mdwtab}
\usepackage{bbm}
\usepackage{bm}
\usepackage{dsfont}
\usepackage{enumerate}
\usepackage{stfloats}
\usepackage{algorithm}

\usepackage{flushend}

\newlength{\bibitemsep}
\newlength{\bibparskip}
\let\oldthebibliography\thebibliography
\renewcommand\thebibliography[1]{%
  \oldthebibliography{#1}%
  \setlength{\parskip}{\bibitemsep}%
  \setlength{\itemsep}{\bibparskip}%
}

\newtheorem{theorem}{Theorem}
\newtheorem{lemma}{Lemma}

\usepackage{array}

\usepackage{url}

\def\MA{\textsf{Max-Agree}}
\def\MC{\textsf{Max-Cut}}

\begin{document}
%
\title{Network Clustering via Maximizing Modularity:\\ Approximation Algorithms and Theoretical Limits}


\author{\IEEEauthorblockN{Thang N. Dinh\IEEEauthorrefmark{1}\IEEEauthorrefmark{3},
Xiang Li\IEEEauthorrefmark{2}, and
My T. Thai\IEEEauthorrefmark{2}},\\
\IEEEauthorblockA{\IEEEauthorrefmark{1}Department of Computer Science,
Virginia Commonwealth University, Richmond, VA 23284 USA},\\
\IEEEauthorblockA{\IEEEauthorrefmark{2}Dept. of Comp. \& Info. Sci. \& Eng., University of Florida, Gainesville, FL 32611 USA}\\
\IEEEauthorblockA{\IEEEauthorrefmark{3}Corresponding author: Thang N. Dinh, email: tndinh@vcu.edu}
}


%


\maketitle

\begin{abstract}
Many  social networks  and complex systems are found to be naturally divided into clusters of densely connected nodes, known as community structure (CS). 
 Finding CS is one of fundamental yet challenging topics in network science.  
One of the most popular classes of methods for this problem is to maximize Newman's modularity. However, there is a little understood on how well we can approximate the maximum modularity as well as the implications of finding community structure with provable guarantees. In this paper, we settle definitely the approximability of modularity clustering, proving that approximating the problem within any (multiplicative) positive factor is intractable, unless \textbf{P}~=~\textbf{NP}. Yet we propose the first additive approximation algorithm for modularity clustering with a constant factor. Moreover, we provide a rigorous proof that  a CS with modularity arbitrary close to maximum modularity $Q_{OPT}$  might bear no similarity to the optimal CS of maximum modularity. Thus even when CS with near-optimal modularity are found, other verification methods are needed to confirm the significance of the structure.     
\end{abstract}


%
\IEEEpeerreviewmaketitle

\section{Introduction}
Many complex systems of interest such as the Internet, social, and biological
relations, can be represented as networks consisting a set of \emph{nodes} which
are connected by \emph{edges} between them. Research in a number of academic
fields has uncovered unexpected structural properties of complex networks
including small-world phenomenon \cite{Watts98}, power-law degree distribution, and the existence of \emph{community structure} (CS) \cite{Girvan02} where
nodes are naturally clustered into tightly connected modules, also known as
communities, with only sparser connections between them. Finding this community~structure is a fundamental but challenging problem in the study of network
systems and has not been yet satisfactorily solved, despite the huge effort of a large
interdisciplinary community of scientists working on it over the past  years
\cite{Fortunato08}.


Newman-Girvan's modularity  that measures the ``strength'' of partition of a
network into modules (also called communities or clusters) \cite{Girvan02}  has
rapidly become an essential element of many community detection methods.
 Despite of the known drawbacks \cite{Fortunato07, Good10}, modularity is by far the most used and best known quality function,
 particularly because of its successes in many  social and
 biological networks \cite{Girvan02} and the ability to auto-detect the optimal number of clusters \cite{ Ruan09, Shakarian13}.
One can search for community structure  by looking for the
divisions of a network that have positive, and preferably large, values of the
modularity. This is the underlying ``\emph{assumption}''  for numerous optimization methods that
find communities in the network via maximizing modularity (aka \emph{modularity clustering}) as surveyed
in~\cite{Fortunato08}. However, there is a little understood on the complexity and approximability of modularity clustering besides its \textbf{NP}-completeness \cite{Brandes08, Dinh15_mod} and \textbf{APX}-hardness \cite{Dasgupta13}.
The approximability of modularity clustering in general graphs remains an open question.

This paper focuses on understanding theoretical aspects of CSs with \emph{near-optimal} modularity. Let $\mathcal C^*$ be a CS with maximum modularity value and let $Q_{OPT}$ be the modularity value of $\mathcal C^*$. Given  $0 < \rho < 1$,  polynomial-time algorithms that can find CSs with modularity at least $\rho Q_{OPT}$ are called (multiplicative) \emph{approximation algorithms}; and $\rho$ is called (multiplicative) approximation factor. 
 Given the \textbf{NP}-completeness of modularity clustering, we are left with two choices: designing heuristics which provides no performance guarantee (like the vast major modularity clustering works) or designing approximation algorithms which can guarantee  near-optimal modularity.  

We seek the answers to the following questions: how well we can approximate the maximum modularity, i.e., for what values of $\rho$ there exist $\rho$-approximation algorithms for modularity clustering? Moreover, do CSs with near-optimal modularity bear similarity to $\mathcal C^*$, the ultimate target of all modularity clustering algorithms?
Our contributions (answers to the above questions)  are as follows.
\begin{itemize}
  \item  We prove that there is \emph{no approximation algorithm with any factor $\rho > 0$ for modularity clustering}, unless \textbf{P} = \textbf{NP},  therefore definitively settling the approximation complexity of the problem. We prove this intractability results for both weighted networks and unweighted networks (with the allowance of multiple edges.)

  \item On the bright side, we propose the \emph{first additive approximation algorithm} that find a community structure with modularity at least   $Q_{OPT}-2(1-\kappa)$ for $\kappa=0.766$. The proposed algorithm also  provides better quality solutions  comparing to the-state-of-the-art modularity clustering methods.
  
    \item We provide rigorous proof that CSs with near-optimal modularity  might be completely different from $\mathcal C^*$, the CS with maximum modularity $Q_{OPT}$. This holds no matter how close the modularity value to $Q_{OPT}$ is. Thus adopters of modularity clustering should carefully employ other verification methods even when they found CSs with modularity values that are extremely close to the optimal ones.  
\end{itemize}   

\textbf{Related work.} A vast amount of methods to find community structure is
surveyed in \cite{Fortunato08}.  Brandes et al. proves the \textbf{NP}-completeness for 
 modularity clustering, the first hardness result for this problem. 
 The problem stands NP-hard even for trees \cite{Dinh15_mod}. 
 DasGupta et
al.  show that  modularity clustering  is APX-hard, i.e., there exists a constant $c> 1$ so that there is no (multiplicative) $c$-approximation for  modularity clustering unless \textbf{P}=\textbf{NP} \cite{Dasgupta13}. In this paper, we  show a much stronger result that the inapproximability holds for \emph{all} $c >1$.  

Modularity has several known drawbacks. Fortunato and Barthelemy
\cite{Fortunato07} has shown  the resolution limit, i.e., modularity clustering
methods fail to detect communities smaller than a scale, the resolution limit
only appears when the network is substantially large \cite{Lancichinetti09}.
Another drawback is modularity's highly degenerate energy landscape
\cite{Good10}, which may lead to very different 
partitions with  equally high modularity.  However, for small and medium networks of several thousand nodes,
the Louvain method \cite{Blondel08} to optimize modularity  is among the best
algorithms according to the  LFR benchmark  \cite{Lancichinetti09}. The method is also adopted in products such as LinkedIn InMap or Gephi.

While approximation algorithms for modularity clustering in special classes of graphs are proposed 
 for scale-free networks\cite{Dinh12,Dinh13adaptive} and  $d$-regular graphs \cite{Dasgupta13}, no such algorithms for general graphs are known.

\textbf{Organization}. We present terminologies in Section~\ref{sec:prem}. The inapproximability of modularity clustering in weighted and unweighted networks is presented in Section \ref{sec:approx}.  We present  the first additive approximation algorithm for modularity clustering in Section~\ref{sec:algo}. Section \ref{sec:sep} illustrates that the optimality of modularity does not correlate to the similarity between the detected CS and the maximum modularity CS. Section~\ref{sec:exp} presents computational results  and we conclude in Section~\ref{sec:con}.  

 \vspace{-0.12in} 
\section{Preliminaries}
\vspace{-0.12in}
\label{sec:prem}
We consider a network represented as an undirected graph $G=(V,E)$\ consisting
of $n=|V|$ vertices and $m=|E|$ edges. The \emph{adjacency matrix} of $G$ is
denoted by $\bm A= \left(  A_{ij}\right)$, where $A_{ij}$\ is the weight of edge
$(i, j)$ and $A_{ij} =  0$ if $(i, j)\notin E$. We also denote the (weighted)
degree of vertex $i$, the total weights of  edges incident at $i$, by
$\textsf{deg}(i)$ or, in short, $d_i$.
 
\emph{Community structure} (CS)  is a division of the vertices in $V$ into a collection
of disjoint subsets of vertices $\mathcal{C}=\left\{C_1, C_2,\ldots,C_l
\right\}$ that the union gives back $V$. Especially, the \emph{number of communities $l$ is not known as a prior}. Each subset $C_i \subseteq V$ is called
a \emph{community} (or module) and we wish to have more edges connecting vertices in the
same communities than edges that connect vertices in different communities.  In this paper, we shall use the terms community structure and \emph{clustering} interchangeably. 

The \emph{modularity} \cite{Newman06} of $\mathcal{C}$ is defined as
\begin{eqnarray}
\label{def:modularity}
Q(\mathcal{C}) = \frac{1}{2M}\displaystyle\sum_{i,j \in V} \left(A_{ij} - \frac{d_i d_j}{2M}\right)\delta_{ij}
\end{eqnarray}
where $d_i$ and $d_j$ are \emph{degree} of nodes $i$ and $j$, respectively; $M$ is the total edge weights; and the element $\delta_{ij}$ of the
\emph{membership matrix} $\bm \delta$ is defined as \[\delta_{ij} = \begin{cases} 1, &
\mbox{if } i \mbox{ and } j \mbox{ are in the same community} \\ 0, &
\mbox{otherwise}.\end{cases}.\] The modularity values can be either positive or
negative and it is believed that the higher (positive) modularity values indicate stronger community
structure. The \emph{ modularity clustering problem} asks to find a division
which maximizes the modularity value.

 Let $\bm B$ be the \emph{modularity matrix} \cite{Newman06} with entries  
\[
B_{ij} = A_{ij} - \frac{d_i d_j}{2M}. \text{ We have } Q(\mathcal{C}) = \frac{1}{2M}\sum_{i,j}B_{ij} \delta_{ij}.
\] 
%
%
Alternatively the modularity can also be defined as
\begin{eqnarray}
\label{def:modularity2}
Q(\mathcal{C}) = \displaystyle\sum_{t=1}^l \left(\frac{E(C_t)}{M} - 
\frac{vol(C_t)^2}{4M^2}\right),
\end{eqnarray}
where $E(C_t)$  is the total weight of the edges  inside $C_t$ and
 $vol(C_t) = \sum_{v \in C_t} d_v$ is the
\emph{volume} of  $C_t$.

\section{Multiplicative Approx. Algorithm}
\label{sec:approx}
A major thrust in optimization is to develop approximation algorithms of which one can theoretically prove the performance bound.  Designing approximation algorithms is, however, very challenging. Thus, it is desirable to know for what values of $\rho$, there exist $\rho$-approximation algorithms.  This section gives a negative answer to the existence of approximation algorithms for modularity clustering with any (multiplicative) factor $\rho >0$, unless \textbf{P} $=$ \textbf{NP}.

We show the inapproximability result for weighted networks via a gap-producing redution from the PARTITION problem in subsection \ref{subsec:inapprox}. Ignoring the weights doesn't make the problem any easier to approximate, as we shall show in subsection~\ref{subsec:unweighted} that the same inapproximability hold for unweighted networks.

Our proofs for both cases use the fact that we can approximate modularity clustering if and only if we can approximate the problem of partitioning the network into two communities to maximize modularity. Then we show that the later problem cannot be approximated within any finite factor.

\subsection{\ Inapproximability in Weighted Graphs}
\label{subsec:inapprox}


\begin{figure}[htb]
\centering 
\includegraphics[width=0.4\textwidth]{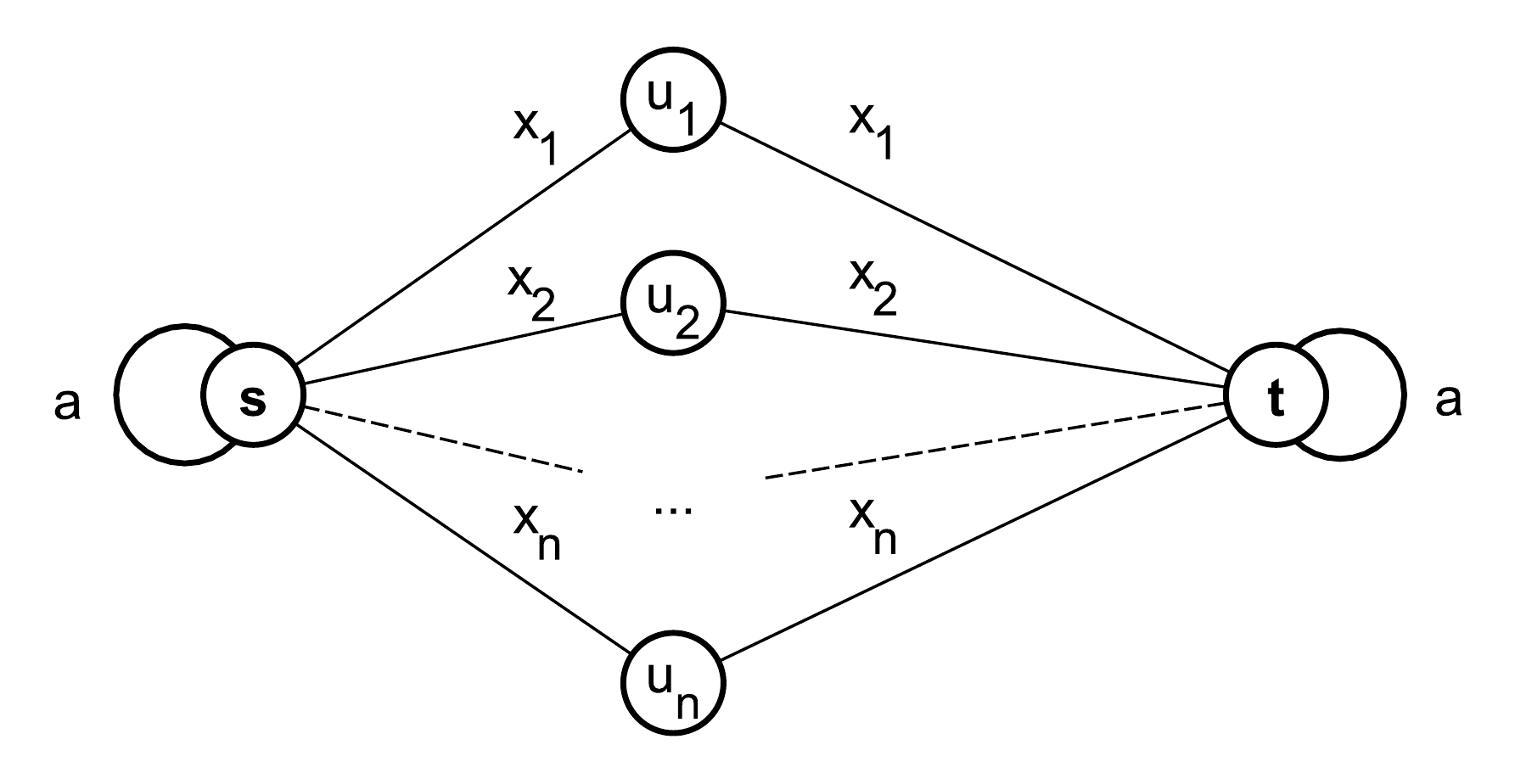} 
\caption{\small Gap-producing reduction from PARTITION to modularity clustering. There exists a community structure of positive modularity if and only if we can divide the integers $x_1,\ldots,x_n$ into two halves with equal sum.} 
\label{fig:reduction}
\end{figure}

\begin{theorem} \label{theo:main}
 For any $\rho >0$, there is no polynomial-time algorithm to find a community structure with a modularity value at least $\rho Q_{OPT}$, unless \textbf{P}$=$\textbf{NP}. Here $Q_{OPT}$ denotes the maximum modularity value among all possible divisions of the network into communities. 
\end{theorem} 
\begin{IEEEproof}
We present a \emph{gap-producing} reduction \cite{Arora09_comp}  that maps an instance $\Phi$ of the following problem

PARTITION: Given integers $x_1, x_2, \ldots, x_n$, can we divide the integers into two halves with equal sum?

to a graph $\tilde G=(\tilde V, \tilde E)$ such that
\begin{itemize}
  \item If $\Phi$ is an YES instance, i.e., we can divide $x_i$ into two halves with equal sum, then $Q_{OPT}(\tilde G) > 0$.
  \item If $\Phi$  is a NO instance, then $Q_{OPT}(\tilde G) = 0$.
\end{itemize}

\emph{Reduction}: The graph $\tilde G$ is shown in Fig. \ref{fig:reduction}. $\tilde G$ consists of two special nodes $s$ and $t$ and $n$ middle nodes $u_1, u_2,\ldots, u_n$. Each $u_i$ is connected to both $s$ and $t$ with edges of weights $x_i$.  Let $K = \frac{1}{2} \sum_{t=1}^n x_t$. Both $s$ and $t$ have self-loops of weights $a = \frac{1}{8K+2} $. 
The total weights of edges in $\tilde G$ is \[
\tilde M = 2\sum_{t=1}^n x_t + 2a  = 4K + 2a.
\]

This reduction establishes the \textbf{NP}-hardness of distinguish graphs having a community structure of positive modularity from those having none. An approximation algorithm with a guarantee $\rho>0$ or better, will find a community structure of modularity at least $\rho Q_{OPT}(\tilde G) > 0$, when given a graph from the first class. Thus, it can distinguish the two classes of graphs, leading to a contradiction to the \textbf{NP}-hardness of PARTITION \cite{Garey90}.  

($\bm \rightarrow$) If $\Phi$ is an YES instance, there exists a partition of $\{ 1, 2, \ldots,n \}$ into disjoint subsets $S_1$ and $S_2$  such that 
\[
\sum_{i \in S_1} x_i = \sum_{j \in S_2} x_j = \frac{1}{2} \sum_{t=1}^n x_t=K,
\]
 Consider a CS $\tilde {\mathcal C}$ in $\tilde G$ that consists of two communities $C_1 = \{ s \} \cup \{ u_i | i \in S_1 \}$ and $C_2 = \{ t \} \cup \{ u_j | j \in S_2 \}$. We have $vol(C_1) = vol(C_2)= \tilde M$. From (\ref{def:modularity2}), the modularity value of $\tilde C$ is 
\begin{align*}
Q(\tilde{\mathcal  C})= \frac{2K+2a}{\tilde M} - \frac{2\tilde M^2}{4 \tilde M^2}=\frac{a}{\tilde M} > 0
\end{align*}

Thus $Q_{OPT} \geq Q_{\tilde{\mathcal  C}} > 0$.

($\bm \leftarrow$) If $\Phi$ is a NO instance, we prove by contradiction that $Q_{OPT}=0$. Assume otherwise $Q_{OPT} > 0$. Let $Q_2$ denote the maximum modularity value among all  partitions of $\tilde G$ into (at most) \emph{two communities}. It is known from \cite{Dinh12} that 
\[ Q_2 \geq \frac{1}{2} Q_{OPT}. \]
Thus there exists a community $\hat{\mathcal C}$ of modularity value $Q_2 \geq \frac{1}{2} Q_{OPT} > 0$ such that $\hat{\mathcal C}$ has exactly two communities, say $\hat C_1$ and $\hat C_2$. Let $\delta(\hat C_1)$ be the total weights of edges crossing between $\hat C_1$ and $\hat C_2$. We have  
\[
Q_2 =  \frac{\tilde M - \delta(\hat C_1)}{\tilde M} - \frac{vol(\hat C_1)^2 + vol(\hat C_2)^2}{4\tilde M^2}.
\]
Substitute $2\tilde M = vol(\hat C_1) + vol(\hat C_2)$ and simplify
\begin{align}
\label{eq:q2}
Q_2 &= \frac{1}{4\tilde M^2}\left( 2vol(\hat C_1)vol(\hat C_2) - 4\tilde M \delta(\hat C_1) \right)\\
\nonumber &= \frac{vol(\hat C_1)vol(\hat C_2)}{2\tilde M^2}\Big(1- \Big[\frac{\delta(\hat C_1)}{vol(\hat C_1)} + \frac{\delta(\hat C_1)}{vol(\hat C_2)} \Big]  \Big)
\end{align}		  	
Since $Q_2 > 0$, we have
\begin{align}
\label{eq:ncut}
\frac{\delta(\hat C_1)}{vol(\hat C_1)} + \frac{\delta(\hat C_1)}{vol(\hat C_2)}  < 1. 
\end{align}

We show that  $s$ and $t$ cannot be in the same community. Otherwise, assume $s$ and $t$ belong to $\hat C_1$, then $\hat  C_2$ contains only nodes from $\{ u_1, u_2,\ldots,u_n\}$. Thus \[ vol(\hat C_2) = \delta(\hat C_1) = 2\sum_{u_j \in \hat C_2} x_j. \] It follows that $\frac{\delta(\hat C_1)}{vol(\hat C_2)} = 1$, which contradicts (\ref{eq:ncut}). 

Since $s$ and $t$ are in different communities, whether we assign $u_i$ into $\hat C_1$ or $\hat C_2$, it always contributes to $\delta(\hat C_1)$ an amount $x_i$. Therefore 
\[
	\delta(\hat C_1) = \sum_{t=1}^n x_t = 2 K = \frac{1}{2}\tilde M - a.
\]
Since $\Phi$ is a NO instance, the integrality of $x_i$ leads to 
\[vol(\hat C_1) - vol(\hat C_2) = 2 \left( \sum_{u_i \in C_1} x_i - \sum_{u_j \in C_2} x_j  \right) \geq 2.\]
Moreover, $a = \frac{1}{8K+2} < \frac{1}{2\hat M}$. Thus we have
\begin{align*}
&\frac{\delta(\hat C_1)}{vol(\hat C_1)} + \frac{\delta(\hat C_1)}{vol(\hat C_2)}
\geq \delta(\hat C_1) \left( \frac{1}{\tilde M -1} + \frac{1}{\tilde M + 1}\right)\\
= &\frac{ (\frac{1}{2}\tilde M - a) 2\tilde M} {\tilde M^2 - 1} 
>  \frac{\tilde M^2 - 2 \frac{1}{2\tilde M}\tilde M} {\tilde M^2 - 1} = 1,
\end{align*} 
which contradicts (\ref{eq:ncut}).

Hence if $\Phi$ is a NO instance, then $Q_{OPT}=0$.
\end{IEEEproof}

\subsection{\ Inapproximability in Unweighted Graphs}
\label{subsec:unweighted}
This section shows that it is \textbf{NP}-hard to decide whether one can divide  an \emph{unweighted} graph into  communities with (strictly) positive  modularity score. Thus approximating modularity clustering is \textbf{NP}-hard for any positive approximation factor.  Our proof reduces from the unweighted \MC{} problem, which is \textbf{NP}-hard even for 3-regular graphs \cite{Matula90}. Our reduction is explicit and can be used to generate hard instances for modularity clustering problem, as shown in Section \ref{sec:exp}.

\textbf{Remark that} one can replace weighted edges with multiple parallel edges in the reduction in Theorem \ref{theo:main} to get a reduction for unweighted graphs. However, such an approach does not yield a polynomial-time reduction, since  instances of PARTITION can have items with \emph{exponentially large weights}.

\begin{theorem}
\label{theo:unweighted}
	Approximating modularity clustering within any positive factor in unweighted graphs (with the allowance of multiple edges) is \textbf{NP}-hard.
\end{theorem}
\begin{IEEEproof}
We reduce from an  instance $\Psi$ of the \MC{} problem ``\emph{whether an undirected unweighted graph $G=(V, E)$ has a subset $S \subseteq V$ of the vertices such that the size of the cut $\delta(S) = \{ (u, v) \in E \ | \ u \in S, v \notin S\}$ is at least $k$?}'' to a graph $G'=(V', E')$ such that
\begin{itemize}
  \item If the answer to $\Psi$ is YES, i.e., there exists a cut $S$ with $\delta(S)\geq k$, then $Q_{OPT}(G') > 0$.
  \item If the answer to $\Psi$  is NO,  $Q_{OPT}(G') = 0$.
\end{itemize}

 \begin{figure}[hbt!]
  \centering     
    \includegraphics[width=0.4\textwidth]{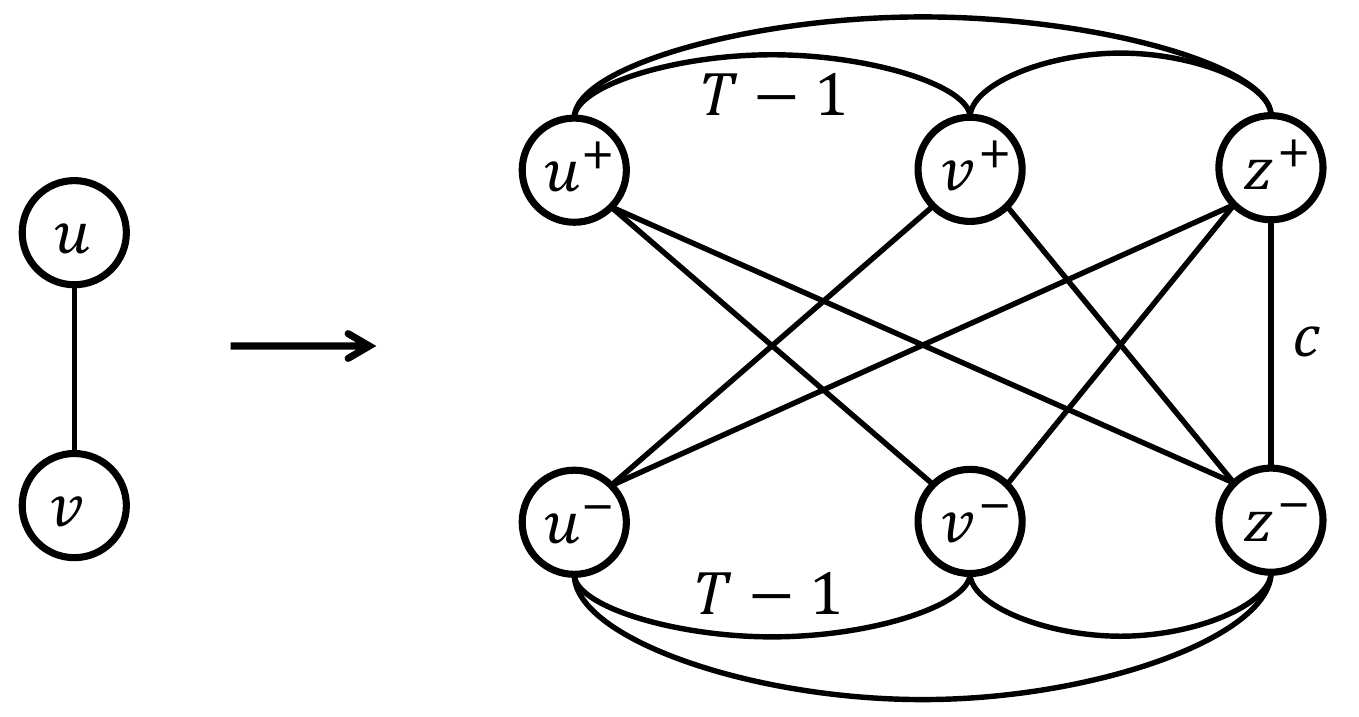}
\caption{\small Reduction for a sample network with an edge connecting two nodes. The multiplicity of edges in the right network is $T=n^4$ unless otherwise noted.}
  \label{fig:unweight}    
\end{figure}

 Using the same arguments in the proof of Theorem~\ref{theo:main}, the above reduction leads to the \textbf{NP}-hardness of approximating modularity clustering within any positive finite factor in unweighted graphs. 

Our reduction is similar to the reduction from \MC{} in \cite{Matula90}. An example is given in Fig. \ref{fig:unweight}.
 For each vertex $v \in V$, we add two vertices $v^+$ and $v^-$ into $V'$. Also we add two special vertices $z^+$ and $z^-$ into $V'$. Thus $V' = \{ v^+, v^-\ |\ v \in V \} \cup \{ z^+, z^-\}$.   Next choose a large integer constant $T=n^4$, where $n=|V|$. We connect vertices in $G'$ in the following orders:
\begin{itemize}
  \item For each edge $(u, v) \in E$, connect $u^+$ to $v^+$ and $u^-$ to $v^-$, each using $T-1$ parallel edges.
  \item There are no edges between $u^+$ and $u^-$ for all $u\in V$. Connect $z^+$ to $z^-$ using $c$ parallel edges, where $c = 4k - 2m -1$ (and $m = |E|$).
  \item Connect the remaining pairs of vertices, each using $T$ parallel edges.
\end{itemize}  
\emph{Feasibility of Reduction.} Obviously, the reduction has a polynomial size. Denote by $n'$ and $m'$ the number of vertices and edges in $G'$, respectively. We have 
\[n' = 2n + 2 \text{ and } 
m' =  2n(n+1) T - 2m + c.
\] 
We also need to verify that $c \geq 0$. By \cite{Vitanyi81}, we can always find in $G$ a cut of size at least $\frac{m}{2} + 2$, thus we can distinguish trivial instances of \MC{} with $k \leq \frac{m}{2} + 2$ from the rest in a polynomial time. For non-trivial instances of \MC{}, i.e., $k > \frac{m}{2} + 2$ we have $4k -2m - 1 >4(\frac{m}{2}+2) - 2m -1 >0$.

($\bm \rightarrow$) If $\Psi$ is an YES instance, there exists a cut $(S \subseteq V, \bar{S} = V\setminus S)$ satisfying $\delta_G(S) \geq k$. Let $S^+ = \{ v^+|\ v \in S\}, \bar{S}^+ = \{ v^+|\ v \notin S\}, S^- = \{ v^-|\ v \in S\},$ and $\bar{S}^- = \{ v^-|\ v \notin S\}$. 
Construct a CS $\mathcal C=\{ C_1, C_2\}$ of $G'$ in which 
\[ C_1 = S^+ \cup \bar{S}^- \cup \{z^+\}, C_2 = S^- \cup \bar{S}^+ \cup \{z^-\}.\] 

We will prove that $Q(\mathcal C) > 0$. By Eq.~(\ref{eq:q2}), 
\begin{align}
\label{eq:mod2}
Q(\mathcal C) = \frac{1}{4 m'^2}\left( 2vol( C_1)vol( C_2) - 4 m'  \delta_{G'}(C_1) \right)
\end{align}
Observe that $d_{v^+} = d_{v^-}=2nT - d_v, \forall v \in V$ and both communities $C_1$ and $C_2$ either contains $v^+$ or $v^-$ but not both. The same observation holds for the vertices $z^+$ and $z^-$ that have degrees $2nT + c$. Thus
\begin{align}
\label{eq:vol}
vol(C_1) = vol(C_2) = m'.
\end{align}

To compute $\delta_{G'}(C_1)$, we recall that the nodes in $C_1$ connect to those in $C_2$, each with $T$ parallel edges with the exceptions of the following pairs:
\begin{itemize}
  \item $2\delta_G(S)$ pairs of nodes between $(S^+, \bar{S}^+)$ and $(S^-, \bar{S}^-)$, each connected with $T-1$ parallel edges
  \item $z^+$ connects to $z^-$ with only $c$ parallel edges.  
\end{itemize} 
Hence, we have 
\begin{align}
\nonumber \delta_{G'}(C_1) &= n(n+1) T - 2\delta_G(S) + c\\
\label{eq:cutopt}				 &\leq n(n+1) T - 2k + c.
				\end{align} 
Substitute Eqs. (\ref{eq:vol}) and (\ref{eq:cutopt}) into  (\ref{eq:mod2}), we have
\begin{align*}
&Q(\mathcal C) = \frac{1}{4 m'^2}( 2m'^2 - 4 m'  \delta_{G'}(C_1) )
= \frac{  m' - 2 \delta_{G'}(C_1)}{2m'}\\
&\geq \frac{1}{2m'}\Big( 2n(n+1) T - 2m + c\\ 
&\quad\quad	- 2 n(n+1) T + 4k - 2c\Big) = \frac{1}{2m'} > 0.
\end{align*}
 
Thus $Q_{OPT} \geq Q_{\mathcal  C} > 0$.

($\bm \leftarrow$) If $\Psi$ is a NO instance, we prove by contradiction that $Q_{OPT}=0$. Assume otherwise $Q_{OPT} > 0$. Let $Q_2$ denote the maximum modularity value among all  partitions of $G'$ into (at most) \emph{two communities} and $\mathcal C=\{ C_1, C_2\}$ be a community structure of $G'$ with the modularity value $Q_2 \geq \frac{1}{2} Q_{OPT} > 0$ \cite{Dinh12}. We will show that $Q_2 \leq 0$, hence, a contradiction. Assume that $y=|C_1| \leq |C_2|$, consider the  following two cases:
  
\textbf{Case $y < n + 1$:}  Since $d_{v^+} = d_{v^-}=2nT - d_v, \forall v \in V$ and  $d_{z^+} = d_{z^-}=2nT + c$, we have
\[
	vol(C_1) \leq 2nTy + 2c 
\]
Since  $vol(C_1) + vol(C_2) = 2m'$, it follows that
\[
vol(C_1)vol(C_2) \leq (2nTy + 2c)(2m' - (2nTy + 2c)).  
\]
Moreover, using the same arguments that leads to Eq.~\ref{eq:cutopt}, we have 
\[
\delta_{G'}(C_1) \geq y(2n+2-y)T-yT=yT(2n+1-y).
\]
Here the factor $yT$ arises from the fact that there are at most $y$ pairs of $(v^+, v^-)$ that across $C_1$ and $C_2$. 

Thus we obtain from (\ref{eq:mod2}) the following inequality
 \begin{align*}
 &Q(\mathcal C) = 
 \frac{1}{4 m'^2}\left( 2vol( C_1)vol( C_2) - 4 m'  \delta_{G'}(C_1) \right)\\
 &\leq \frac{1}{2 m'^2}\Big( (2nTy + 2c)(2m' - (2nTy + 2c))\\
&\quad\quad\quad\quad  - 2 m'yT(2n+1-y)   \Big).
\end{align*}
After some algebra and applying the inequalities $y~\leq~n$ and $c \leq 2n^2$, we obtain
\begin{align*}
Q(\mathcal C)\leq \frac{2T^2ny}{m'^2}\left( -(n+1 -y) +\frac{O(n^3)}{T}\right) < 0.
 \end{align*}

\textbf{Case $y=|C_1| = |C_2| = n + 1$:}
We bound $\delta_{G'}(C_1)$ by considering two sub-cases:
\begin{itemize}
  \item 
 If there is some $v\in V$ such that $v^+, v^- \in C_1$ or $z^+, z^- \in C_1$, then $\delta_{G'}(C_1) \geq (n+1)(n+1)T - nT -(n+1)(n+1)$  
 \item
 Otherwise, all pairs $v^+$ and $v^-$ (as well as $z^+$ and $z^-$) are in different sides of the cut $C_1$. Thus $C_1$ induces in $G$ a cut $S \subseteq V$. Then $\delta_{G'}(C_1) \geq n(n+1)T - 2 \delta_{G'}(S)+c \geq n(n+1)T-2(k-1)+c$, as $\delta(S) < k$. 
 \end{itemize} 
As $n(n+1)T + T-(n+1)^2 \geq n(n+1)T-2(k-1)+c$, it holds for the both cases that \[
\delta_{G'}(C_1)  \geq n(n+1)T-2(k-1)+c.\]
Since \[
vol( C_1)vol( C_2) \leq m'^2,
\]
using Eq. (\ref{eq:mod2}), we obtain 
 \begin{align*}
 &Q(\mathcal C) \leq 
 \frac{1}{4 m'^2}\left( 2m'^2 - 4 m'\left(n(n+1)T-2(k-1)+c\right)   \right)\\
 &\leq \frac{1}{2m'} \Big( 2n(n+1)T -2m + c\\
 &\quad\quad - 2n(n+1)T + 4k-4 -2c\Big) = \frac{-3}{2m'} < 0.
\end{align*}
Thus if $\Psi$ is a NO instance, then $Q_{OPT}=0$.
\end{IEEEproof}

\section{Additive Approx. Algorithm}
\label{sec:algo}
We propose \emph{the first {additive approximation} algorithm} that find a community structure $\mathcal C$  satisfying the following performance guarantee
\begin{align}
\label{eq:addmul}
Q(\mathcal C) \geq  Q_{OPT}  - 2(1-\kappa),
\end{align}
where $\kappa = 0.766$. The algorithm is based on rounding a semidefinite programm, similar to that in \cite{Charikar05} for the \MA{} problem.

First, we formulate modularity clustering as a vector programming. Let $e_j \in \mathbb R^n$ be  the unit vector with $1$ in the $i^\text{th}$ coordinate and $0$s everywhere else. Let $x_i \in \{ e_1, e_2,\ldots,e_n \}$ be the variable that indicates the community of vertex $i$, i.e., if $x_i = e_j$ then vertex $i$ belongs to community $j$. The vector programming is as follows.
\begin{align}
\label{sdp:obj}\max\quad &\frac{1}{2M}\sum_{i,j} B_{ij}\quad x_i \cdot x_j \\
 & x_i \in \{ e_1, e_2, \ldots, e_n\}\quad \forall i,
 \end{align}
where $(\cdot)$ denotes the inner product (or dot product).
 
We  relax the constraint $x_i \in \{ e_1, e_2, \ldots,e_n\}$ to get a semidefinite program (SDP) with new constraints
\begin{align}
\label{sdp:norm}\quad &x_i\cdot x_i = 1\quad \forall i\\
 	 &x_i\cdot x_j \geq 0\quad \forall i \neq j\\
\label{sdp:range} 	 & x_i \in \mathbb R^n \quad \forall i.
\end{align}



One of the reason that modularity clustering resists approximation approaches such as semidefinite rounding is that the matrix $\bm B$ contains both negative and nonnegative entries. Indeed, all entries in $\bm B$ sum up to zero \cite{Newman06}. To overcome this, we add a fixed amount $\frac{W}{2M}$ to the objective of SDP, where 
\begin{align*}
&	W = \sum_{(i, j) \in \bm B^+} B_{ij} = |\sum_{(i, j) \in \bm B^-} B_{ij} | \text{ with }\\
&\bm B^+ = \{ (i, j)\ |\ B_{ij} \geq 0 \} \text{ and } \bm B^-= \{ (i, j)\ |\ B_{ij} < 0 \}.
\end{align*} 

The new objective is then  
\begin{align}
\nonumber	&\frac{1}{2M}\Big( \sum_{i,j} B_{ij} x_i \cdot x_j - \sum_{(i, j) \in \bm B^-} B_{ij} \Big)\\
\nonumber	=&\frac{1}{2M}\Big(\sum_{ (i,j) \in \bm B^+}B_{ij} x_i \cdot x_j  + 
\sum_{(i, j) \in \bm B^-} B_{ij} (x_i \cdot x_j - 1) \Big)\\
\nonumber =&\frac{1}{2M}\big(\sum_{ (i,j) \in \bm B^+}B_{ij} x_i \cdot x_j +	 
	 \sum_{(i, j) \in \bm B^-}-B_{ij} (1 - x_i \cdot x_j)\big).
\end{align}
Note that all of coefficients in the new objective are \emph{nonnegative}. Thus we transform the modularity clustering problem to an SDP of the \MA{} problem \cite{Charikar05} which can be solved using the rounding procedure in \cite{Charikar05}. Our additive approximation algorithm can be summarized as follows.
\begin{algorithm}
  \caption{SDP to Maximize Modularity  (SDPM)}
  \label{alg:sdpm}
  \begin{algorithmic}[1]
   \State Solve the SDP relaxation in (\ref{sdp:obj}) and (\ref{sdp:norm})-(\ref{sdp:range})
   \State  Choose $k$ random hyperplanes, and use projection to divide the set of vertices into $2^k$ clusters.
   \State Return the better clustering $\mathcal C$ of $k=2$ and $k=3$.
  \end{algorithmic}
\end{algorithm}

Since all coefficients in the new objective are positive and the fixed factor $\frac{W}{2M}$ does not affect the solution of SDP. We can apply Theorem 3 in \cite{Charikar05} to obtain
\begin{align}
\label{eq:rel} Q_G(\mathcal C) + \frac{W}{2M}\geq \kappa \Big(Q_{OPT} + \frac{W}{2M}\Big),
\end{align}
where $\kappa=0.766$ is the  approximation factor for the generalized \MA{} problem \cite{Charikar05}.

Since $\frac{W}{2M} < 1$ and $Q_{OPT} < 1$,  we can  simplify (\ref{eq:rel}) to yield the following theorem.

\begin{theorem}
\label{theo:app}
Given graph $G$, there is a polynomial-time algorithm that finds a community structure $\mathcal C$ of $G$ satisfying
 \[
 	 Q_G(\mathcal C) > \kappa Q_{OPT} - (1-\kappa),
 \]
 and
 \[
 	 Q_G(\mathcal C) >  Q_{OPT} - 2(1-\kappa).
 \] 
 where $\kappa = 0.766$.
\end{theorem}

Apparently, the higher $\kappa$ the better the performance guarantee.  Any improvement on the approximation factor for the generalized \MA{} problem will immediately lead to the improvement in the approximation factor for modularity clustering.
\section{Do Small Gaps Guarantee Similarity?}
\label{sec:sep}
Given $0 < a < b < 1$ and an arbitrary graph $G$, we show how to construct a ``structurally equivalent'' graph $G'$ of $G$ in which  community structures have modularity values  between $a$ and $b$. Multiple implications of this finding include:
\begin{itemize}

  \item There are graphs of any size that have clustering with extremely small modularity (e.g. by choosing $a$ and $b$ close to zero.) This gives additional light into why it is hard to distinguish between graphs having no community structure with positive modularity and the others (Section \ref{subsec:inapprox}.)
    \item There are graphs of any size that all ``reasonable'' clustering of the network yields modularity values in range $(a(1-\epsilon), a)$ for arbitrary small $\epsilon>0$ and any $0 < a < 1 -\epsilon$. Thus even we find a  CS with modularity at least $(1-\epsilon)Q_{OPT}$ or $Q_{OPT} -\epsilon$, the obtained CS can be completely different from $\mathcal C^*$, the  maximum modularity CS.
\end{itemize}

Therefore, the presence of high modularity clusters neither indicates the presence of community structure nor how easy it is to detect such a structure if it exists.

We present our construction which consists of two transformations, namely $\alpha$-transformation and $(\tau, k)$-transformation.

\textbf{$\alpha$-transformation}: An $\alpha$-transformation with $0 < \alpha \leq 1$ maps each graph $G=(V, E)$ with an ``equivalent'' graph $G'=T_{\alpha}(G)$ and maps (one-to-one correspondence) each CS $\mathcal C$ of $G$ to a CS $\mathcal C'$ of $G'$ that satisfies \[ Q_{G'}(\mathcal C') = \alpha Q_G(\mathcal C), \] where $Q_{G'}(\mathcal C')$ and $Q_G(\mathcal C)$ denote the modularity  of $\mathcal C'$ in $G'$ and $\mathcal C$ in $G$, respectively. 

\emph{Construction:}  $G'$ also has $V$ as the set of vertices. The weighted adjacency matrix $A'$ of $G'$ is defined as
\begin{align}
\label{eq:defa}
A'_{ij} = A_{ij} + \frac{1 -\alpha}{\alpha} \frac{d_i d_j}{2M}.
\end{align}

We show in the following lemma that the same community induced by $\mathcal C$ in $G'$ has modularity scaled down by a fraction $\alpha$.
\begin{lemma}
Given a community structure $\mathcal C$ of $G$, the CS $\mathcal C'$ induced by $\mathcal C$ in $G'=T_\alpha(G)$ satisfies
\[ Q_{G'}(\mathcal C') = \alpha Q_G(\mathcal C). \]
\end{lemma}
\begin{IEEEproof}
Let $\delta_{ij} = 1$ if $i$ and $j$ are in the same community in $C$ and $\delta_{ij}=0$ otherwise. By definition 
\[
  Q_{G'}(\mathcal C') = \frac{1}{2M'}\sum_{i, j}\left( A_{ij}' - \frac{d_i' d_j'}{2M'}\right)\delta_{ij},
\] 
where $M', d_i'$, and $d_j'$ are the total edge weights, weighted degree of $i$, and weighted degree of $j$ in $G'$, respectively.

We have
\begin{align}
\nonumber d_i' &= \sum_{j \in V} A_{ij}' = \sum_{j \in V} \left( A_{ij} + \frac{1-\alpha}{\alpha} \frac{d_i d_j}{2M} \right)\\
\label{eq:di}	&= \sum_{j \in V} A_{ij} + \frac{1 -\alpha}{\alpha} d_i \sum_{j \in V} d_j/(2M) = \frac{1}{\alpha} d_i.
\end{align}
Moreover,
\begin{align}
\label{eq:M}
M' &= \frac{1}{2}\sum_{i \in V} d_i' = \frac{1}{2\alpha} \sum_{i \in V} d_i = \frac{1}{\alpha} M.
\end{align} 
From (\ref{eq:defa}), (\ref{eq:di}), and (\ref{eq:M}), we have
\begin{align*}
Q_{G'}(\mathcal C') &= \frac{\alpha}{2M}\sum_{i, j}\left( A_{ij} + \frac{1 -\alpha}{\alpha} \frac{d_i d_j}{2M} - \frac{d_i d_j }{2M \alpha}\right)\delta_{ij}\\
		   &= \frac{\alpha}{2M}\sum_{i, j}\left( A_{ij}  - \frac{d_i d_j }{2M }\right)\delta_{ij}
		   = \alpha Q_G(\mathcal C).\ 
\end{align*}
\end{IEEEproof}

\textbf{$(\tau, k)$-transformation}: A $(\tau,k)$-transformation with $0 < \tau < 1$ and $k \in \mathsf{Z}^+$ maps a graph $G=(V, E)$ with a graph $G'=T_{\tau, k}(G)$ and maps each community structure $C$ in $G$ to a community structure $C'$ in $G'$ that satisfies
\[
Q_{G'}({\mathcal C'}) = \tau + (1-\tau - \epsilon) Q_G(\mathcal C),
\]
where $\epsilon = \frac{(1-\sqrt{\tau})^2}{k}$.

\emph{Construction}: The set of vertices $V'$ is obtained by adding to $V$  $k$ isolated vertices $n+1, n+2,\ldots,n+k$.
Let $\beta = \frac{1}{\sqrt \tau} -1$, i.e., $\tau = 1/(1+\beta)^2$. We attach loops of weight $\frac{\beta}{2}d_i$ to  vertices $1 \leq i \leq n$ and  loops of weight $\frac{\beta(\beta+1)}{k} M$ to $n+1, \ldots,n+k$. Thus the weighted adjacency matrix $A'$ of $G'$ is as follows.
\begin{align}
 A_{ij}' = \begin{cases}
 				A_{ij} &  1 \leq i \neq j \leq n \\
 				\frac{\beta}{2} d_i & 1 \leq i = j \leq n \\
 				\frac{1}{k}\beta(\beta+1) M & i = j > n  \\
 				0 & \text{ otherwise }.
 		  \end{cases}
\end{align}
CS $\mathcal C'$ of $G'$ is obtained from $\mathcal C$ by adding $k$ singleton communities that contains only one node from $\{ n+1,\ldots,n+k \}$.

\begin{lemma}
Given a community structure $\mathcal C$ of $G$, the community structure $\mathcal C'$ induced by $\mathcal C$ in $G'=T_{\tau,k}(G)$ satisfies
\[
Q_{G'}({\mathcal C'}) = \tau + (1-\tau - \epsilon) Q_G(\mathcal C),
\]
where $\epsilon = \frac{(1-\sqrt{\tau})^2}{k}$.
\end{lemma}
\begin{IEEEproof}
 Since a loop contribute twice to the degree, we have
\begin{align}
\label{eq:di2}
 d_i' = \sum_{j \neq i} A_{ij} + 2 \frac{\beta}{2} d_i = (1 + \beta) d_i,
\end{align}  
and
\begin{align}
\label{eq:dn12}
 d_{n+l}' =\frac{2}{k} \beta (\beta+1) M, l=1..k.
\end{align}  
Therefore
\begin{align}
\label{eq:M2}
\nonumber M' &= \frac{1}{2} \sum_{i \in V'} d_i' = \frac{1}{2}\left( \sum_{i \in V} d_i' +  k \frac{2}{k}\beta (\beta+1) M \right)\\
   &=   (1 + \beta)M + \beta (\beta+1)M  =  (\beta+1)^2 M.
\end{align}
We have
\begin{align*}
 Q_{G'}(\mathcal C') = \frac{1}{2M'}\sum_{i, j \in V} \left(A_{ij}' - \frac{d_i' d_j'}{2M'}  \right)\delta_{ij}\\
 		 +  \sum_{l=1}^k \left(\frac{\beta(\beta+1)M}{kM'} - \frac{d_{n+l}'^2}{4M'^2}\right)\delta_{n+l, n+l}.
\end{align*}
Substitute (\ref{eq:di2}), (\ref{eq:dn12}), and (\ref{eq:M2}) into the above equation
\begin{align*}
Q_{G'}(\mathcal C') = \frac{1}{2M (\beta+1)^2}\sum_{i, j \in V} \left(A_{ij} - \frac{d_i d_j}{2M}  \right)\delta_{ij}\\
 +\frac{\sum_{i \in V}\frac{\beta}{2} d_i}{M'}	+  \left( \frac{\beta}{\beta+1} - \frac{\beta^2}{k(1+\beta)^2}\right)\\
 	= \frac{1}{(\beta+1)^2}Q_G(\mathcal C) + 1 - \frac{1}{(\beta+1)^2} - \frac{\beta^2}{k(\beta+1)^2}\\
 	= \tau Q_G(\mathcal C) + (1 - \tau - \epsilon).
\end{align*}
This yields the proof.
\end{IEEEproof}

Now we can combine the two transformations to ``engineer'' the modularity values into any desirable range $(a, b)$. 
\begin{theorem}
Given a graph $G$, applying an $\alpha$-transformation on $G$, followed by a $(\tau, k)$-transformation yields a graph $\tilde G$ and a mapping from  each community structure $C$ of $G$ to a community structure $\tilde C$ of $\tilde G$ that satisfies
\[
  Q_{\tilde G}(\tilde{ \mathcal C})=   \tau \alpha Q_G(\mathcal C) + (1 - \tau - \epsilon),
\]  
where $\epsilon = \frac{(1-\sqrt{\tau})^2}{k}$.
\end{theorem}

Since $-1/2 < Q_{ G}({\mathcal  C}) < 1$ \cite{Dinh12}, setting $\tau = 1 - (\frac{2}{3}a + \frac{1}{3}b)$ and $\alpha=\frac{2}{3}(b-a)$ ensures that $a < Q_{\tilde G}(\tilde{\mathcal  C}) < b$ for any $0 < a < b < 1$.

\section{Computational Results}
\label{sec:exp}
\begin{table}[hbt]
  \centering
  \caption{\small Order and size of network instances}
    \begin{tabular}{ccrr}
    \addlinespace
    \toprule
     ID & Name  & $n$ &\quad $m$ \\
    \midrule
    1     & Zachary's karate club & 34    & 78 \\
    2     & Dolphin's social network      &  62      &  159\\
    3     & Les Miserables      &    77   &  254 \\
    4     & Books about US politics      &  105    & 441  \\
    5     & American College Football     &    115  & 613  \\    
    6     & Electronic Circuit (s838)      &   512   &  819  \\
    \bottomrule
    \end{tabular}%
  \label{tab:sum}%
\end{table}%


We compare the modularity values of the most popular algorithms in the literature
\cite{Girvan02, Newman06, Agarwal08} to that of the SDP rounding in Alg. \ref{alg:sdpm} (SDPM). Also, we include the state of the art, the Louvain (aka Blondel's) method, \cite{Blondel08}. Since Blondel is a stochastic algorithm, we repeat the algorithm 20 times and report the best modularity value found. 
The optimal modularity values are reported in \cite{Cafieri10}. For solving SDP, we use SDTP3 solver \cite{Tutuncu03} and repeat the rounding process 1000 times and pick the best result. All algorithms are run on a PC with a Core i7-3770 processor and 16GB RAM. 
 
\subsection{\ Real-world networks} 
  We perform the experiments on the standard datasets for community structure identification \cite{Agarwal08, Cafieri10}, consisting of real-world networks. The datasets' names   together with their sizes are  listed in Table \ref{tab:sum}.  

\begin{table}[t!]
  \centering
  \caption{\small Comparing modularity obtained by different methods CNM (fast-greedy) \cite{Clauset04}, EIG \cite{Newman06}, Louvain  \cite{Blondel08}, SDPM, the semidefinite rounding in this paper, and the optimal modularity values OPT \cite{Cafieri10}.}
    \begin{tabular}{ccrrrrrrr}
    \addlinespace
    \toprule
    ID  &   CNM &\quad EIG & Louvain & SDPM  & OPT\\
    \midrule
    1        & 0.235 & 0.393 &\textbf{ 0.420}   &0.419 &  0.420\\
    2        & 0.402 & 0.491     & \textbf{0.529}   & 0.526 & 0.529\\
    3        & 0.453 & 0.532     & \textbf{0.560}   &\textbf{0.560} &  0.560\\
    4       & 0.452  & 0.467 &\textbf{ 0.527}   & \textbf{0.527} & 0.527\\
    5       & 0.491 & 0.488     & \textbf{0.605}   & \textbf{0.605} & 0.605\\      
    6       & 0.803 & 0.736     & 0.796   & -  & 0.819\\ 
    \bottomrule
    \end{tabular}%
  \label{tab:mod}%
\end{table}%

The results are reported in Table~\ref{tab:mod}. The SDP method finds community structures with \emph{maximum modularity} (\emph{optimal}) values. Our SDPM method has  high running-time and space-complexity. It ran out of memory for the largest test case of 512 nodes and  819 edges. However, it  not only approximates the maximum modularity much better than the (worst-case) theoretical performance guarantee, Theorem \ref{theo:app}, but also is among the highest quality modularity clustering methods.
%
 \begin{figure}[hbt]
  \centering    
    \includegraphics[width=0.36\textwidth]{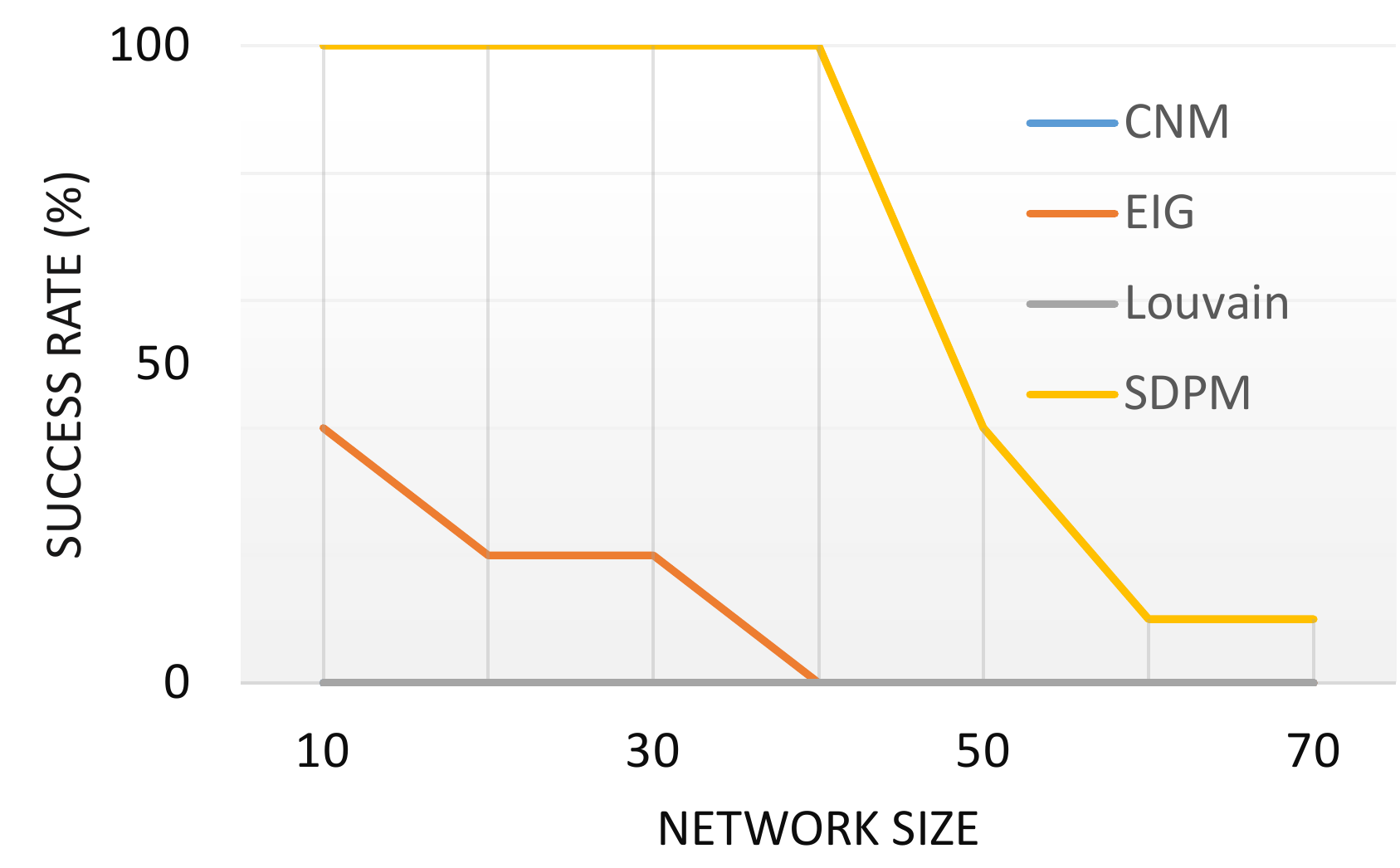}
  \caption{\small Success rate of finding CSs with positive modularity values in the hard instances.}
  \label{fig:hardnp}  
\end{figure}
  
\subsection{\ Hard Instances via \MC{} reduction}
To validate the effectiveness of   modularity clustering methods, we generate hard instances of modularity clustering via the reduction from \MC{} problem in the proof of Theorem \ref{theo:unweighted}. The advantages of this type of test includes:  1) Generated networks are small but yet challenging to solve and 2) Optimal solutions and objective (modularity) are known. This contrasts other test generators such as LFR \cite{Lancichinetti09} that often come with planted community structure but not (guaranteed) optimal solutions.

We generate the tests following the below steps:
\begin{itemize}
  \item Generate a random (Erd\H{o}s-R\'{e}yni) network $G$.
  \item Find the exact size $k$ of the \MC{} in $G$ using the Biq Mac solver \cite{Rendl10}.
  \item Construct a network $G'$ from the instance $\langle G, k\rangle$ of \MC{} using the reduction in Theorem \ref{theo:unweighted}.
  \item Run modularity maximization methods on $G'$. A method passes a test if it can find a community structure with a strictly positive modularity value.  
\end{itemize}

We vary network sizes between 10 to 70, increasing by 10 and repeat the test five times for each network size. The number of times each method passes the test  are shown in Fig~\ref{fig:hardnp}. Our  SDPM algorithm clearly has much higher success rate than the rest. It passes all the tests of size up to 40. The only method that manages to pass some of the tests is the Eigenvector-based method (EIG) \cite{Newman06}. EIG passes the tests of sizes 10, twice and sizes 20 and 30, once. These tests illustrates the excellent capability of the SDP rounding methods for hard-instances of the modularity clustering problem.

\section{Conclusion}
\label{sec:con}
In this paper, we settle the question on the approximability of modularity clustering. We show that there is no (multiplicative) approximation algorithm with any factor $\rho >0$, unless \textbf{P} = \textbf{NP}. However, we show that there is an additive approximation algorithm that find community structure with modularity at least $\kappa Q_{OPT} - (1-\kappa)$ with $\kappa =0.766$. Not only modularity is hard to approximate,  but also it is a poor indicator for the existing of community structure. The existing of high modularity clusters
neither indicates the existing of community structure
nor how easy it is to detect such a structure if it exists. 

In the future, it is interesting to investigate additive approximation algorithms for modularity clustering, i.e., algorithms to find CS with modularity at least $Q_{OPT} - c$ for $c>0$. We conjecture that there exists  $c > 0$ that approximating modularity clustering within an additive approximation factor $c$ is \textbf{NP}-hard.

\section{Acknowledgement}
This work is partially supported by NSF CAREER 0953284
and NSF CCF 1422116.
\bibliographystyle{IEEEtran}
\bibliography{partitioning,complexity,modularity}

\begin{thebibliography}{10}
\providecommand{\url}[1]{#1}
\csname url@samestyle\endcsname
\providecommand{\newblock}{\relax}
\providecommand{\bibinfo}[2]{#2}
\providecommand{\BIBentrySTDinterwordspacing}{\spaceskip=0pt\relax}
\providecommand{\BIBentryALTinterwordstretchfactor}{4}
\providecommand{\BIBentryALTinterwordspacing}{\spaceskip=\fontdimen2\font plus
\BIBentryALTinterwordstretchfactor\fontdimen3\font minus
  \fontdimen4\font\relax}
\providecommand{\BIBforeignlanguage}[2]{{%
\expandafter\ifx\csname l@#1\endcsname\relax
\typeout{** WARNING: IEEEtran.bst: No hyphenation pattern has been}%
\typeout{** loaded for the language `#1'. Using the pattern for}%
\typeout{** the default language instead.}%
\else
\language=\csname l@#1\endcsname
\fi
#2}}
\providecommand{\BIBdecl}{\relax}
\BIBdecl

\bibitem{Watts98}
D.~J. Watts and S.~H. Strogatz, ``{Collective dynamics of 'small-world'
  networks},'' \emph{Nature}, vol. 393, 1998.

\bibitem{Girvan02}
M.~Girvan and M.~E. Newman, ``Community structure in social and biological
  networks.'' \emph{PNAS}, vol.~99, no.~12, 2002.

\bibitem{Fortunato08}
S.~Fortunato and C.~Castellano, ``Community structure in graphs,'' \emph{Ency.
  of Complexity and Sys. Sci.}, 2008.

\bibitem{Fortunato07}
S.~Fortunato and M.~Barthelemy, ``Resolution limit in community detection,''
  \emph{Proceedings of the National Academy of Sciences}, vol. 104, no.~1,
  2007.

\bibitem{Good10}
B.~H. Good, Y.-A. de~Montjoye, and A.~Clauset, ``Performance of modularity
  maximization in practical contexts,'' \emph{Phys. Rev. E}, vol.~81, p.
  046106, Apr 2010.

\bibitem{Ruan09}
J.~Ruan, ``A fully automated method for discovering community structures in
  high dimensional data,'' in \emph{Proc. of the IEEE Int. Conf. on Data Mining
  (ICDM)}, 2009, pp. 968--973.

\bibitem{Shakarian13}
P.~Shakarian, P.~Roos, D.~Callahan, and C.~Kirk, ``Mining for geographically
  disperse communities in social networks by leveraging distance modularity,''
  in \emph{Proc. of the ACM Int. Conf. on Knowledge Discovery and Data Mining
  (KDD)}, 2013, pp. 1402--1409.

\bibitem{Brandes08}
U.~Brandes, D.~Delling, M.~Gaertler, R.~Gorke, M.~Hoefer, Z.~Nikoloski, and
  D.~Wagner, ``On modularity clustering,'' \emph{Knowledge and Data
  Engineering, IEEE Transactions on}, vol.~20, no.~2, 2008.

\bibitem{Dinh15_mod}
T.~N. Dinh and M.~T. Thai, ``Toward optimal community detection: From trees to
  general weighted networks,'' \emph{Internet Mathematics}, vol.~11, no.~3, pp.
  181--200, 2015.

\bibitem{Dasgupta13}
B.~Dasgupta and D.~Desai, ``On the complexity of newman's community finding
  approach for biological and social networks,'' \emph{J. Comput. Syst. Sci.},
  vol.~79, no.~1, pp. 50--67, Feb. 2013.

\bibitem{Lancichinetti09}
A.~Lancichinetti and S.~Fortunato, ``Community detection algorithms: A
  comparative analysis,'' \emph{Phys. Rev. E}, vol.~80, p. 056117, Nov 2009.

\bibitem{Blondel08}
V.~D. Blondel, J.-L. Guillaume, R.~Lambiotte, and E.~Lefebvre, ``{Fast
  unfolding of communities in large networks},'' \emph{Journal of Statistical
  Mechanics: Theory and Experiment}, vol. 2008, no.~10, 2008.

\bibitem{Dinh12}
T.~N. Dinh and M.~T. Thai, ``Community detection in scale-free networks:
  Approximation algorithms for maximizing modularity,'' in \emph{IEEE Journal
  on Selected Areas in Communications}, 2013.

\bibitem{Dinh13adaptive}
T.~N. Dinh, N.~P. Nguyen, and M.~T. Thai, ``An adaptive approximation algorithm
  for community detection in dynamic scale-free networks,'' in
  \emph{Proceedings IEEE INFOCOM}, 2013, pp. 55--59.

\bibitem{Newman06}
M.~E.~J. Newman, ``{Modularity and community structure in networks},''
  \emph{Proceedings of the National Academy of Sciences}, vol. 103, 2006.

\bibitem{Arora09_comp}
S.~Arora and B.~Barak, \emph{Computational Complexity: A Modern Approach},
  1st~ed.\hskip 1em plus 0.5em minus 0.4em\relax New York, NY, USA: Cambridge
  University Press, 2009.

\bibitem{Garey90}
M.~R. Garey and D.~S. Johnson, \emph{Computers and Intractability: A Guide to
  the Theory of NP-Completeness}.\hskip 1em plus 0.5em minus 0.4em\relax New
  York, NY, USA: W. H. Freeman \& Co., 1990.

\bibitem{Matula90}
D.~W. Matula and F.~Shahrokhi, ``Sparsest cuts and bottlenecks in graphs,''
  \emph{Discrete Applied Mathematics}, vol.~27, no. 1–2, pp. 113 -- 123, 1990.

\bibitem{Vitanyi81}
P.~Vitanyi, ``How well can a graph be n-colored?'' \emph{Discrete mathematics},
  vol.~34, no.~1, pp. 69--80, 1981.

\bibitem{Charikar05}
M.~Charikar, V.~Guruswami, and A.~Wirth, ``Clustering with qualitative
  information,'' \emph{Learning Theory, J. of Comput. Syst. Sci.}, vol.~71,
  no.~3, pp. 360 -- 383, 2005.

\bibitem{Agarwal08}
G.~Agarwal and D.~Kempe, ``Modularity-maximizing graph communities via
  mathematical programming,'' \emph{Eur. Phys. J. B}, vol.~66, 2008.

\bibitem{Cafieri10}
D.~Aloise, S.~Cafieri, G.~Caporossi, P.~Hansen, S.~Perron, and L.~Liberti,
  ``Column generation algorithms for exact modularity maximization in
  networks.'' \emph{Phys. Rev. E}, vol.~82, 2010.

\bibitem{Tutuncu03}
R.~H. T\"{u}t\"{u}nc\"{u}, K.~C. Toh, and M.~J. Todd,
  ``\BIBforeignlanguage{English}{Solving semidefinite-quadratic-linear programs
  using {SDPT3}},'' \emph{\BIBforeignlanguage{English}{Mathematical
  Programming}}, vol.~95, no.~2, pp. 189--217, 2003.

\bibitem{Clauset04}
A.~Clauset, M.~E.~J. Newman, and C.~Moore, ``Finding community structure in
  very large networks,'' \emph{Phys. Rev. E}, vol.~70, p. 066111, Dec 2004.

\bibitem{Rendl10}
F.~Rendl, G.~Rinaldi, and A.~Wiegele, ``Solving {M}ax-{C}ut to optimality by
  intersecting semidefinite and polyhedral relaxations,'' \emph{Math.
  Programming}, vol. 121, no.~2, p. 307, 2010.

\end{thebibliography}

\end{document}